\documentclass[prd,aps]{revtex4}
\usepackage{amsmath}

\begin{document}

\title{Small Black Holes on Branes: Is the horizon regular or singular ?}
\author{D. Karasik}
\email{karasik@physics.uc.edu}
\author{C. Sahabandu}
\author{P. Suranyi}
\author{L. C. R. Wijewardhana}
\affiliation{Department  of Physics,
University of Cincinnati, Cincinnati, Ohio}

\begin{abstract}
We investigate the following question:
Consider a small mass,
with $\epsilon$ (the ratio of the Schwarzschild radius and the bulk curvature length)
much smaller than 1, that is confined to the TeV brane
in the Randall-Sundrum I scenario.
Does it form a black hole with
a regular horizon, or a naked singularity?
The metric is expanded in $\epsilon$
and the asymptotic form of the metric is given by the
weak field approximation (linear in the mass).
In first order of $\epsilon$
we show that the iteration of the weak field solution,
which includes only integer powers of the mass,
leads to a solution that has a singular horizon.
We find a solution with a regular horizon but its asymptotic
expansion in the mass also contains half integer powers.
\end{abstract}
\pacs{04.50.+h, 04.70.Bw, 04.70.Dy}
\maketitle

\section{Introduction}
Black holes in theories with extra dimensions have been studied
widely.    Myers and Perry~\cite{myers} found Schwarzschild type
solutions (MPS) in  $D$-dimensional asymptotically flat space.  Black hole
solutions were also found in asymptotically AdS space~\cite{hawking1,birmingham}.
No non-trivial closed form  black hole solutions, other than the
black string solution \cite{hawking2}, which extends in a uniform
manner from the brane into the extra dimension,  have been found
in three-brane theories of the Randall-Sundrum type \cite{rs1,rs2}.
Given that there is
considerable interest surrounding the production of black holes
at accelerators~\cite{giddings}, and in collisions of cosmic
rays \cite{shapere1},
    it is important to develop approximate methods to find
black hole solutions in Randall-Sundrum brane world theories.

   Some initial attempts at finding black hole solutions centered on
deriving the induced metric on the brane by solving the
Hamiltonian constraint conditions~\cite{induced}. Some of the
induced solutions do not arise from  matter distributions confined
to the brane~\cite{kanti}. Linearized solutions about RS
backgrounds~\cite{lin1} as well as numerical
solutions~\cite{num,kudoh} have also been derived.

In a recent paper \cite{us} we have studied the metric around a
small mass that is confined to the TeV brane in Randall-Sundrum
type I scenario (RSI)\cite{rs1}. The configuration is
characterized by two length scales; the Schwarzschild radius
$\mu$, which is related to the mass $\mu=\sqrt{8G_{5}M/(3\pi)}$,
and the curvature length of the bulk $\ell$, which is related to
the cosmological constant and the brane tension. We study the case
$\mu\ll\ell$
where the metric can be expanded in the dimensionless
parameter, $\epsilon=\mu/\ell\ll1$. There are two ways to make
$\epsilon\ll1$: (1) $\mu\rightarrow0$ while $\ell$ is finite. This
limit is just the original Randall-Sundrum scenario in the absence of matter.
Keeping $\ell$ finite and expanding in $\mu$ (or actually in $M$)
is called the weak field approximation. First order in $M$ is
linearized gravity. Linearized gravity is valid at
distances much larger than the Schwarzschild radius, $r\gg\mu$. (2)
$\ell\rightarrow\infty$ while $\mu$ is finite. This limit is a
five dimensional black hole in an asymptotically flat background.
The metric for this configuration is the Myers and Perry solution
(MPS). We studied the expansion in $\epsilon$ with $\mu$ kept
finite. We call the first order expansion in $\epsilon$ - the $\epsilon$
solution. The $\epsilon$ solution is valid at distances much
smaller than the bulk curvature length $r\ll\ell$.

On one hand, the $\epsilon$ solution is needed to study the horizon
and the thermodynamics of the black hole, for which the linearized
solution is unsuitable.  On the other hand, the $\epsilon$ solution
cannot be fixed uniquely without satisfying certain boundary conditions.
We need the linearized solution to identify the mass and to satisfy
the junction conditions on the Planck brane because the distance
between the branes in RSI is of the same order as the
bulk curvature length $d\sim\ell$.

Since we need both the $\epsilon$ solution and the linearized solution
we look at the region where both solutions are valid,
$\mu\ll r\ll\ell$. In this region we expand the linearized
solution to first order in $\ell^{-1}$ and the $\epsilon$ solution
to first order in $M$. We require that these two expansions will
coincide. Using this method, we are able to incorporate the
asymptotic characteristics into the short-ranged $\epsilon$
solution. In other words, the linearized solution is the
asymptotic boundary condition for the $\epsilon$ solution.

In this paper we study the regularity of the horizon. In section
\ref{sec:epsilon} we summarize the results from \cite{us}. In
section \ref{sec:horizon} we analyze the horizon and formulate the
conditions such that the $\epsilon$ solution will be regular on
the horizon. We also calculate the thermodynamics parameters of
the black hole assuming it is regular. In section \ref{sec:repH}
we represent the $\epsilon$ solution with Legendre functions. We
assume that in the asymptotic region the metric is given by the
linearized solution (linear in the mass). Another reasonable
assumption is that when iterating the linearized solution, the
post linearized solution includes only integer powers of the mass.
This assumption is motivated by the post Newtonian behavior of $4$
dimensional black holes, where the asymptotic solution is an
expansion in integer powers of $G_{4}M/r$. We show that under this
assumption the horizon is singular. In section
\ref{sec:postlinearized} we show that one can construct a regular
solution provided that the post linearized solution includes half
integer powers of the mass. We find it interesting that the first
law of black hole thermodynamics, $dM=T\,dS$, eliminates the terms
of order $M^{3/2}$ from the asymptotic expansion.

\section{The $\epsilon$ solution}
\label{sec:epsilon}
We use the following ansatz for the metric
\begin{equation}
    ds^{2}=a(\rho,\psi)\left[-B(\rho,\psi)dt^{2}
    +\frac{A(\rho,\psi)}{B(\rho,\psi)}d\rho^{2}+
    2V(\rho,\psi)d\rho\,d\psi+\rho^{2}U(\rho,\psi)d\psi^{2}
    +\rho^{2}\sin^{2}\psi\,d\Omega^{2}_{2}\right]~,
    \label{ansatz}
\end{equation}
where $d\Omega^{2}_{2}$ is the metric on the unit $2$-sphere. The
conformal factor is taken from RSI as
$a(\rho,\psi)=(1-\epsilon\rho|\cos\psi|)^{-2}$. The TeV brane is
located at $\psi=\pi/2$ as can be read off of the conformal factor. We
work only in the interval $0\leq\psi\leq\pi/2$, and assume $Z_{2}$
symmetry about the brane. The point mass is located at $\rho=0$.
The $\epsilon$ solution is just MPS solution with $\epsilon$ corrections
\begin{subequations}
\label{emetric}
\begin{eqnarray}
    B(\rho,\psi)&=&1-\frac{1}{\rho^{2}}-\epsilon B_{1}(\rho,\psi)~,\\
    A(\rho,\psi)&=&1+\epsilon A_{1}(\rho,\psi)~,\\
    U(\rho,\psi)&=&1+\epsilon U_{1}(\rho,\psi)~,\\
    V(\rho,\psi)&=&\epsilon V_{1}(\rho,\psi)~.
\end{eqnarray}
\end{subequations}
We have rescaled the coordinates with $\mu$ such that $\rho=1$
corresponds to the MPS horizon at the Schwarzschild radius.
The gauge is partially fixed by the coefficient of $d\Omega^{2}_{2}$.
There is still a gauge freedom associated with the coordinate transformation
$\rho\rightarrow\rho(1-\epsilon F)$, $\psi\rightarrow\psi+\epsilon F \tan\psi$.

The $\epsilon$ solution can be written in terms
of the gauge function $F(\rho,\psi)$ and
the wave function $H(\rho,\psi)$. The solution is
\footnote{This solution is obtained from the solution in \cite{us}
using redefinition of $A\rightarrow A/B$ and redefining the gauge function
$F\rightarrow F-6H_{0}+(2\rho-\rho^{-1})H_{0}'$.}
\begin{subequations}
\label{finalmetric}
\begin{eqnarray}
    B_{1}&=&-\frac{2}{\rho^{2}}F(\rho,\psi)~,
    \label{gtt}\\
    A_{1}&=&8\rho(\rho^{2}-1)\cos\psi-12 H(\rho,\psi)
       -4\tan\psi H_{,\psi}(\rho,\psi)
    +2F(\rho,\psi)+2\rho F_{,\rho}(\rho,\psi)~,
    \label{grhorho}\\
    U_{1}&=&
    \frac{2\tan^{2}\psi}{\rho}\left[6\rho H(\rho,\psi)
    -(2\rho^{2}-1)H_{,\rho}(\rho,\psi)
    -\rho F(\rho,\psi)-\rho\cot\psi F_{,\psi}(\rho,\psi)\right]~,
    \label{gpsipsi}\\
    V_{1}&=&
    -2\rho^{2}(2\rho^{2}-1)\sin\psi
    +2\rho^{2}\tan\psi H_{,\rho}(\rho,\psi)
    \nonumber\\& &
    +\frac{\rho(2\rho^{2}-1)\tan^{2}\psi H_{,\psi}(\rho,\psi)}{\rho^{2}-1}
    -\rho^{2}\tan\psi F_{,\rho}(\rho,\psi)+\frac{\rho^{3}F_{,\psi}(\rho,\psi)}
    {\rho^{2}-1}~.
    \label{grhopsi}
\end{eqnarray}
\end{subequations}
The wave function $H(\rho,\psi)$ satisfies the differential equation
\begin{equation}
    (\rho^{2}-1)\left(H_{,\rho\rho}-\frac{1}{\rho}H_{,\rho}\right)+H_{,\psi\psi}
    +2\frac{\cos^{2}\psi+1}{\sin\psi\cos\psi}H_{,\psi}=0~.
    \label{Hequation}
\end{equation}

{\it Junction conditions on the brane.} Israel junction conditions
\cite{junction} on the brane are simply
$B_{1,\psi}(\rho,\pi/2)=0$, $A_{1,\psi}(\rho,\pi/2)=0$,
$V_{1}(\rho,\pi/2)=0$, and $U_{1}(\rho,\pi/2)<\infty$. These
conditions imply the following
\begin{subequations}
\label{junction}
\begin{eqnarray}
        & &F_{,\psi}(\rho,\pi/2)=0~,\label{F,psi}\\
        & & \left.\left(3H+\tan\psi H_{,\psi}
    \right)_{,\psi}\right|_{\psi=\pi/2}=-2\rho(\rho^{2}-1)~,\label{H3}\\
        & &H_{,\psi}(\rho,\pi/2)=0~,\label{H1}\\
        & & F(\rho,\pi/2)=6H(\rho,\pi/2)
    -\frac{2\rho^{2}-1}{\rho}H_{,\rho}(\rho,\pi/2)~.\label{Fbrane}
\end{eqnarray}
\end{subequations}

{\it Asymptotic boundary conditions.} As mentioned earlier, we
take the linearized solution as an asymptotic boundary condition.
As a result, the asymptotic form of the wave function is
\begin{eqnarray}
    H(\rho,\psi)&=&\delta_{0}+\frac{\rho^{3}}{16\sin^{3}\psi}(4\psi-\sin4\psi)
    +\frac{\rho}{64\sin^{3}\psi}(4\sin2\psi-8\psi\cos2\psi
    -12\psi+3\sin4\psi)\nonumber\\
    & &+\frac{1}{1024\rho\sin^{3}\psi}(1024\delta
    +8\psi+32\psi\cos2\psi+12\psi\cos4\psi-16\sin2\psi-5\sin4\psi)
    +{\cal O}(\rho^{-(1+2\alpha)})~,
    \label{Hasymp}
\end{eqnarray}
where $\delta$ and $\delta_{0}$ are arbitrary constants.
The asymptotic form of the gauge function is
\begin{equation}
    F(\rho,\psi)=\frac{\rho}{2}\left(\cos\psi+\frac{\psi}{\sin\psi}
    \right)+6\delta_{0}+\frac{F_{2}(\psi)}{\rho}
    +{\cal O}(\rho^{-(1+2\alpha)})~,
    \label{Fasymp}
\end{equation}
where $F_{2}(\pi/2)=8\delta-5\pi/64$ and $F''_{2}(\pi/2)=47\pi/64-24\delta$.

In the functions (\ref{Hasymp}) and (\ref{Fasymp}) we have
neglected corrections of order $\rho^{-(1+2\alpha)}$ where
$\alpha>0$. The reason is that such terms will contribute to order
$M^{1+\alpha}$, which is beyond the linearized solution.

A reasonable assumption is that if one will go beyond the
linearized solution, only integer powers of $M$ will appear in the
metric. This assumption will force the functions
$H(\rho,\psi)$ and $F(\rho,\psi)$ to be
odd functions of $\rho$, therefore one must set $\delta_{0}=0$
and the neglected terms are of order $\rho^{-3}$.

\section{Analyzing the horizon}
\label{sec:horizon}
The configuration is called a black hole if there exists a Killing horizon, i.e.
a null surface with a Killing field normal to the surface.
The ansatz (\ref{ansatz}) is static, and the static Killing vector
becomes null on the surface $B=0$. The static Killing vector is
$\xi^{\mu}=\delta^{\mu}_{0}$, it is normalized such that asymptotically $\xi^{2}=-1$.

A vector normal to the surface $B=0$ must be proportional to $n_{\mu}=B_{,\mu}$.
If the surface $B=0$ is null, then the normal vector must be null on $B=0$.
\begin{equation}
    \left.n^{\mu}n_{\mu}\right|_{B=0}=\frac{(B_{,\psi})^{2}}{\rho^{2}a U}=0~.
\end{equation}
Therefore, the derivative $B_{,\psi}$ must vanish on the surface $B=0$.
This implies that the surface $B(\rho,\psi)=0$ is actually defined as
$\rho=\rho_{H}=\text{constant}$. The ansatz for $B(\rho,\psi)$ is
\begin{equation}
    B(\rho,\psi)=\left(1-\frac{\rho_{H}^{2}}{\rho^{2}}\right)b(\rho,\psi)~,
    \label{Bansatz}
\end{equation}
where $b(\rho,\psi)$ does not vanish at $\rho=\rho_{H}$,
but it is regular such that
$\lim_{\rho\rightarrow\rho_{H}}(\rho-\rho_{H})b(\rho,\psi)=0$.

The surface gravity, $\kappa$, for the ansatz (\ref{ansatz}) is
defined as
\begin{equation}
    \kappa^{2}=-\left.\frac{1}{2}g^{\mu\nu}g^{\lambda\sigma}
    \xi_{\mu;\lambda}\xi_{\nu;\sigma}\right|_{B=0}=\frac{(B_{,\rho})^{2}}{4A}~.
    \label{surfacegravity}
\end{equation}
According to a theorem by R\'{a}cz and Wald \cite{wald}, if the surface gravity
is not constant on the horizon, the horizon is singular.
This can be verified in the
ansatz (\ref{ansatz}) by calculating the Kretchmann scalar around the horizon.
Provided that the function $B(\rho,\psi)$ is given by Eq.(\ref{Bansatz}), the
Kretchmann scalar can be expanded in powers of $B$
\begin{equation}
    R^{\mu\nu\lambda\sigma}R_{\mu\nu\lambda\sigma}=\frac{1}{B(\rho,\psi)}
    \frac{32(\kappa_{,\psi})^{2}}{\rho^{2}a^{2}U}
    +\text{regular terms}~.
    \label{Khorizon}
\end{equation}
Clearly, at any point where $\kappa_{,\psi}\neq0$,
the surface $B=0$ is singular. 
If the surface gravity is not constant but vanishing at some points on the
horizon, these points will be regular.
An ansatz for the function $A(\rho,\psi)$, which is compatible with
Eqs.(\ref{Bansatz}, \ref{surfacegravity}), is
\begin{equation}
    A(\rho,\psi)=\frac{b(\rho_{H},\psi)^{2}}{\rho_{H}^{2}\kappa^{2}}
    +\left(1-\frac{\rho_{H}^{2}}{\rho^{2}}\right)\alpha(\rho,\psi)~,
    \label{Aansatz}
\end{equation}
where $\lim_{\rho\rightarrow\rho_{H}}(\rho-\rho_{H})\alpha(\rho,\psi)=0$.

{\it $\epsilon$ expansion on the horizon.} Take the ansatz
(\ref{ansatz}) together with Eqs.(\ref{Bansatz}, \ref{Aansatz})
and expand in $\epsilon$. Zero order in $\epsilon$ should be MPS
solution, therefore, the following $\epsilon$ expansions should be
used
\begin{subequations}
\label{ehorizon}
\begin{eqnarray}
        & & b(\rho,\psi)=1+\epsilon b_{1}(\rho,\psi)~,\label{eb1}\\
    & & \alpha(\rho,\psi)=\epsilon a_{1}(\rho,\psi)~,\label{ea1}\\
    & & \rho_{H}=1+\epsilon\zeta~,\label{defzeta} \\
    & & \kappa=1+\epsilon\chi~.\label{defchi}
\end{eqnarray}
\end{subequations}
Comparing Eqs.(\ref{ehorizon}) with Eqs.(\ref{emetric}), one
can solve for $b_{1}$ and $a_{1}$
\begin{eqnarray}
    b_{1}(\rho,\psi)&=&\frac{2\zeta-\rho^{2}B_{1}(\rho,\psi)}{\rho^{2}-1}~,
    \label{b1}\\
    a_{1}(\rho,\psi)&=&\frac{\rho^{2}\left[A_{1}(\rho,\psi)
    -2b_{1}(1,\psi)+2\zeta+2\chi\right]}{\rho^{2}-1}~.
    \label{a1}
\end{eqnarray}
The expansion in $\epsilon$ on the horizon is possible only if
the functions $b_{1}$ and $a_{1}$ are finite at $\rho=1$.
Therefore, the numerators in Eqs.(\ref{b1}, \ref{a1}) must vanish at $\rho=1$,
and one can calculate $\zeta$, $b_{1}(1,\psi)$, and $\chi$
\begin{eqnarray}
    \zeta&=&\lim_{\rho\rightarrow1}\frac{1}{2}B_{1}(\rho,\psi)~,
    \label{zeta}\\
    b_{1}(1,\psi)&=&-\lim_{\rho\rightarrow1}
    \frac{\left(\rho^{2}B_{1}(\rho,\psi)\right)_{,\rho}}{2\rho}~,
    \label{b11}\\
    \chi&=&\lim_{\rho\rightarrow1}\frac{1}{2}A_{1}(\rho,\psi)
    -b_{1}(1,\psi)-\zeta~.
    \label{chi}
\end{eqnarray}
Both $\zeta$ and $\chi$ are constants, therefore Eqs.(\ref{zeta}, \ref{chi}) put
restrictions on the solution (\ref{finalmetric}). Equation (\ref{zeta}) implies that
\begin{equation}
    \zeta=-F(1,\psi)=\text{constant}\;\;\Rightarrow\; F_{,\psi}(1,\psi)=0~.
    \label{zeta2}
\end{equation}
Substitute the solution
(\ref{finalmetric}) in Eq.(\ref{chi}) to calculate the correction
to the surface gravity
\begin{equation}
    \chi=2\left(3H(1,\psi)+\tan\psi H_{,\psi}(1,\psi)\right)=\text{constant}~.
    \label{chi2}
\end{equation}
So, the combination $3H(1,\psi)+\tan\psi H_{,\psi}(1,\psi)$ must be constant.

Another constraint comes from Eq.(\ref{grhopsi}). The function $V_{1}$ must
be finite at $\rho=1$ (just like $b_{1}$ and $a_{1}$). This implies that
\begin{equation}
    \lim_{\rho\rightarrow1}(\rho^{2}-1)V_{1}(\rho,\psi)
    =\tan^{2}\psi H_{,\psi}(1,\psi)+F_{,\psi}(1,\psi)=0~.
    \label{V1}
\end{equation}
Since $F_{,\psi}(1,\psi)=0$ we deduce that
$H_{,\psi}(1,\psi)=0$. The functions $H$ and $F$ are constants on the horizon
and using Eq.(\ref{Fbrane}) can be evaluated as
\begin{equation}
    H(1,\psi)=H(1,\pi/2)\;\;,F(1,\psi)=F(1,\pi/2)=6H(1,\pi/2)-H_{,\rho}(1,\pi/2)~.
    \label{H1const}
\end{equation}

To summarize, if the $\epsilon$ expansion is valid
on the horizon and the horizon is regular then
\begin{subequations}
\label{regularity}
\begin{eqnarray}
        & & F_{,\psi}(1,\psi)=0~,\label{F1}\\
    & & H_{,\psi}(1,\psi)=0~,\label{H1psi}\\
    & & \rho_{H}=1+\epsilon\zeta=1+\epsilon[-6H(1,\pi/2)+H_{,\rho}(1,\pi/2)]~,\label{rhoH} \\
    & & \kappa=1+\epsilon\chi=1+6\epsilon H(1,\pi/2)~.\label{sg}
\end{eqnarray}
\end{subequations}

{\it Thermodynamics.} If the horizon is regular then one can talk
about thermodynamics of the black hole. The zeroth law of black
hole thermodynamics states that the surface gravity is constant on
the horizon \cite{wald}. The temperature, $T$, associated with the
black hole is
\begin{equation}
    \frac{1}{T}\equiv\frac{2\pi\mu}{\kappa}
    =2\pi\mu\left(1-6\epsilon H(1,\pi/2)\right)~.
    \label{T}
\end{equation}
One should remember that the physical dimension of the surface gravity is
$\text{length}^{-1}$ therefore it is rescaled with $\mu^{-1}$.
The entropy of the black hole is related to the area of the horizon
\begin{equation}
    S\equiv\frac{A_{H}}{4G_{5}}=\frac{2\mu^{3}}{4G_{5}}
    \int_{0}^{\pi/2}d\psi\int_{0}^{\pi}d\theta
    \int_{0}^{2\pi}d\phi\left.\sqrt{g_{\psi\psi}g_{\theta\theta}g_{\phi\phi}}
    \right|_{\rho=1+\epsilon\zeta}=
    \frac{2\pi\mu^{3}\rho_{H}^{3}}{G_{5}}\int_{0}^{\pi/2}d\psi\,\sin^{2}\psi
    \frac{1+\epsilon/2U_{1}(1,\psi)}{(1-\epsilon\cos\psi)^{3}}~.
    \label{horizonarea}
\end{equation}
Using the solution (\ref{finalmetric}) and the constraints
(\ref{H1const}, \ref{regularity}), one can verify that on the
horizon $U_{1}(1,\psi)=2\tan^{2}\psi\left[H_{,\rho}(1,\pi/2)-H_{,\rho}(1,\psi)\right]$.
The entropy is
\begin{equation}
    S=\frac{2\pi^{2}\mu^{3}}{4G_{5}}\left[1+\epsilon\left\{
    3\zeta+\frac{4}{\pi}+\frac{4}{\pi}\int_{0}^{\pi/2}d\psi\,\sin^{2}\psi
    \tan^{2}\psi\left[H_{,\rho}(1,\pi/2)-H_{,\rho}(1,\psi)\right]\right\}\right]~.
    \label{entropy}
\end{equation}
The integral in Eq.(\ref{entropy}) can be simplified using integration by parts as
\begin{eqnarray}
    & &\int_{0}^{\pi/2}d\psi\,\sin^{2}\psi
    \tan^{2}\psi\left[H_{,\rho}(1,\pi/2)-H_{,\rho}(1,\psi)\right]
    =\int_{0}^{\pi/2}d\psi\,\frac{\sin\psi}{\cos^{2}\psi}
    \sin^{3}\psi\left[H_{,\rho}(1,\pi/2)-H_{,\rho}(1,\psi)\right]\nonumber\\
    & &=\left[\frac{\sin^{3}\psi}{\cos\psi}
    \left\{H_{,\rho}(1,\pi/2)-H_{,\rho}(1,\psi)\right\}\right]^{\pi/2}_{0}
    -\int_{0}^{\pi/2}\frac{d\psi}{\cos\psi}\left\{
    \sin^{3}\psi\left[H_{,\rho}(1,\pi/2)-H_{,\rho}(1,\psi)\right]\right\}_{,\psi}
    \nonumber\\
    & &=H_{,\rho\psi}(1,\pi/2)-\frac{3\pi}{4}H_{,\rho}(1,\pi/2)
    +\int_{0}^{\pi/2}\frac{d\psi}{\cos\psi}\left[
    \sin^{3}\psi H_{,\rho}(1,\psi)\right]_{,\psi}~,
    \label{simpintegral}
\end{eqnarray}
where in the last step we have used L'Hopital's rule to evaluate
the boundary term at $\psi=\pi/2$. Using Equations
(\ref{simpintegral}) and (\ref{rhoH}), the entropy (\ref{entropy})
can be written as
\begin{equation}
    S=\frac{2\pi^{2}\mu^{3}}{4G_{5}}\left[1+\epsilon\left\{
    -18H(1,\pi/2)+\frac{4}{\pi}\left[1+H_{,\rho\psi}(1,\pi/2)
    +\int_{0}^{\pi/2}\frac{d\psi}{\cos\psi}\left[
    \sin^{3}\psi H_{,\rho}(1,\psi)\right]_{,\psi}\right]\right\}\right]~.
    \label{entropy2}
\end{equation}

The first law of black hole thermodynamics states that
$T^{-1}=\partial S/\partial M$. In all known black hole solutions
the first law is satisfied as a result of Einstein equations. In
the $\epsilon$ solution it is not trivially satisfied, but should
be imposed as a boundary condition. The mass appears in the
entropy only through $\mu=\sqrt{8G_{5}M/(3\pi)}$ and
$\epsilon=\mu/\ell$, so
\begin{equation}
    \frac{1}{T}=\frac{\partial S}{\partial M}=2\pi\mu\left[1+\epsilon\left\{
    -24H(1,\pi/2)+\frac{16}{3\pi}\left[1+H_{,\rho\psi}(1,\pi/2)
    +\int_{0}^{\pi/2}\frac{d\psi}{\cos\psi}\left[
    \sin^{3}\psi H_{,\rho}(1,\psi)\right]_{,\psi}\right]\right\}\right]~.
    \label{T2}
\end{equation}
Equations (\ref{T2}) and (\ref{T}) must be consistent, therefore,
there is another constraint on the function $H$
\begin{equation}
    H(1,\pi/2)=\frac{8}{27\pi}\left[1+H_{,\rho\psi}(1,\pi/2)
    +\int_{0}^{\pi/2}\frac{d\psi}{\cos\psi}\left[
    \sin^{3}\psi H_{,\rho}(1,\psi)\right]_{,\psi}\right]~.
    \label{T2T1}
\end{equation}

\section{The singular horizon}
\label{sec:repH}
A detailed study of the horizon in the $\epsilon$ solution requires a
specific representation of the function $H(\rho,\psi)$,
which solves the differential equation (\ref{Hequation})
and the boundary conditions (\ref{junction}, \ref{Hasymp}, \ref{regularity}).

A possible representation for the solution
is a combination of Associated Legendre functions
in $\rho$ and Hypergeometric functions in $\psi$
\begin{subequations}
\label{Hlegendre}
\begin{eqnarray}
    H(\rho,\psi)&=&\int\,d\lambda R(\rho;\lambda)\Psi(\psi;\lambda)\label{Hlambda}\\
    R(\rho;\lambda)&=&\rho\sqrt{\rho^{2}-1}\left[a(\lambda)
        Q^{1}_{\frac{\lambda-1}{2}}(2\rho^{2}-1)+b(\lambda)
        P^{1}_{\frac{\lambda-1}{2}}(2\rho^{2}-1)\right]\label{Rlegendre}\\
    \Psi(\psi;\lambda)&=&c(\lambda)\left._{2}F_{1}(\frac{1-\lambda}{2},\frac{1+\lambda}{2},
        \frac{5}{2},\sin^{2}\psi)\right.+\frac{d(\lambda)}{\sin^{3}\psi}\left.
        _{2}F_{1}(\frac{-2-\lambda}{2},\frac{-2+\lambda}{2},
        -\frac{1}{2},\sin^{2}\psi)\right.\label{Psilambda}
\end{eqnarray}
\end{subequations}
A lengthy discussion about the expansion (\ref{Hlegendre}) appears
in \cite{us}. There we assumed that the asymptotic
solution beyond the linearized solution includes only integer
powers of $M$. Therefore, the function $H$ is odd in $\rho$ and is
given by
\begin{eqnarray}
    H(\rho,\psi)&=&\rho\sqrt{\rho^{2}-1}\left[-\frac{2}{3\pi}Q^{1}_{-1/2}(2\rho^{2}-1)
    \frac{3[\sin(2\psi)-2\psi\cos(2\psi)]}{8\sin^{3}\psi}\right.\nonumber\\
    & &\left.+Q^{1}_{1/2}(2\rho^{2}-1)\left(-\frac{1}{3\pi}\frac{3[4\psi-\sin(4\psi)]}
    {32\sin^{3}\psi}
    +\frac{d_{2}}{\sin^{3}\psi}
    -\frac{\pi}{3}\frac{3[8\psi+4\psi\cos(4\psi)-3\sin(4\psi)]}{32\pi^{2}\sin^{3}\psi}
    \right)\right.\nonumber\\
    & &\left.
    +\frac{\pi}{3}\left(P^{1}_{1/2}(2\rho^{2}-1)+\left.\frac{2}{\pi^{2}}
    \frac{\partial Q^{1}_{n-1/2}(2\rho^{2}-1)}{\partial n}\right|_{n=1}\right)
    \frac{3[4\psi-\sin(4\psi)]}{32\sin^{3}\psi}\right.\nonumber\\
    & &\left.
    +\sum_{n=2}^{\infty}a(\lambda=2n)Q^{1}_{n-1/2}(2\rho^{2}-1)
    \frac{3[n\cos(2n\psi)\sin(2\psi)-
    \cos(2\psi)\sin(2n\psi)]}{8n(1-n^{2})\sin^{3}\psi}\right]~.
    \label{Hdiscrete}
\end{eqnarray}
The set $a(\lambda=2n)_{n\geq2}$ is fixed by the junction
condition (\ref{H3})
\begin{equation}
    a(\lambda=2n)_{n\geq2}=\frac{16}{3\pi(\lambda^{2}-4)}~.
    \label{a2n}
\end{equation}
The parameter $d_{2}$ is undetermined.

The problem appears when one tries to apply the conditions
(\ref{chi2}, \ref{regularity}) to the solution (\ref{Hdiscrete}).
\begin{itemize}
\item The associated Legendre functions of the second kind are not
analytic at $\rho=1$
\begin{equation}
    \rho\sqrt{\rho^{2}-1}Q^{1}_{\nu}(2\rho^{2}-1)\sim
    -\frac{1}{2}-\nu(\nu+1)(\rho^{2}-1)\ln|\rho^{2}-1|
    +{\cal O}(\rho^{2}-1)~.
\end{equation}
As a result, one can evaluate the function $H(1,\psi)$ but
not the derivative $H_{,\rho}(1,\psi)$.
\item The surface gravity, (\ref{chi2}), is not constant. The combination
$3H+\tan\psi H_{,\psi}$ is not constant on the horizon
\begin{equation}
    \kappa=1+2\epsilon\left[3H(1,\psi)+\tan\psi H_{,\psi}(1,\psi)\right]
    =1+2\epsilon\left[\frac{2\psi\sin\psi}{\pi}
    +\frac{3\cos\psi}{2\pi}
    -\sum_{n=2}^{\infty}a(\lambda=2n)
    \frac{3\sin(2n\psi)}{4n\sin\psi}\right]
    =1+2\epsilon\sin\psi~.
    \label{notconstant}
\end{equation}
\end{itemize}
As was mentioned earlier, according to the theorem by R\'{a}cz and
Wald \cite{wald}, if the surface gravity is not constant on the
horizon, the horizon is singular. One may notice that on the brane
$\left.\kappa_{,\psi}\right|_{\psi=\pi/2}=0$.
So, according to Eq.(\ref{Khorizon}),
although the horizon is singular, the singularity is naked
from the bulk but it is covered on the brane.

\section{Give up the post linearized assumption}
\label{sec:postlinearized}
We give up the assumption that the asymptotic solution
can be expanded in integer powers of $M$, but still keep the linearized
solution as the asymptotic boundary condition.
According to Eq.(\ref{Hasymp}) we can include terms of order
$\rho^{-(1+2\alpha)}$. It is convenient to work with the Legendre
expansion (\ref{Hlegendre}). The asymptotic behavior of the associated
Legendre functions of the first (second) kind is
$\rho\sqrt{\rho^{2}-1}P(Q)^{1}_{(\lambda-1)/2}(2\rho^{2}-1)\sim\rho^{1+(-)\lambda}$.
Therefore, we can include associated Legendre functions of
the second kind with $\lambda>2$ in the expansion. Actually, we have already
included even $\lambda$ terms in the expansion (\ref{Hdiscrete}).
We first try to include also odd $\lambda$ terms.
\begin{eqnarray}
    H(\rho,\psi)&=&\delta_{0}+\rho\sqrt{\rho^{2}-1}\left[-\frac{2}{3\pi}Q^{1}_{-1/2}(2\rho^{2}-1)
    \frac{3[\sin(2\psi)-2\psi\cos(2\psi)]}{8\sin^{3}\psi}\right.\nonumber\\
    & &\left.+Q^{1}_{1/2}(2\rho^{2}-1)\left(-\frac{1}{3\pi}\frac{3[4\psi-\sin(4\psi)]}
    {32\sin^{3}\psi}
    +\frac{d_{2}}{\sin^{3}\psi}
    -\frac{\pi}{3}\frac{3[8\psi+4\psi\cos(4\psi)-3\sin(4\psi)]}{32\pi^{2}\sin^{3}\psi}
    \right)\right.\nonumber\\
    & &\left.
    +\frac{\pi}{3}\left(P^{1}_{1/2}(2\rho^{2}-1)+\left.\frac{2}{\pi^{2}}
    \frac{\partial Q^{1}_{n-1/2}(2\rho^{2}-1)}{\partial n}\right|_{n=1}\right)
    \frac{3[4\psi-\sin(4\psi)]}{32\sin^{3}\psi}\right.\nonumber\\
    & &\left.
    +\sum_{\lambda=3}^{\infty}a(\lambda)Q^{1}_{(\lambda-1)/2}(2\rho^{2}-1)
    \frac{3[\lambda\cos(\lambda\psi)\sin(2\psi)-2
    \cos(2\psi)\sin(\lambda\psi)]}{2\lambda(4-\lambda^{2})\sin^{3}\psi}\right]~.
    \label{Hevenodd}
\end{eqnarray}
The free parameters are $d_{2}$, $\delta_{0}$, and
$a(\lambda=2k+1)_{k\geq1}$. Our goal is to make the surface gravity constant at
$\rho=1$. Adding the odd $\lambda$ terms does not change the
junction condition
(\ref{H3}) and therefore the coefficients $a(\lambda=2n)$
are the same as in Eq.(\ref{a2n}).
We will try to fix the coefficients $\delta_{0}$, $a(\lambda=2k+1)$ such
that the
combination $3H+\tan\psi H_{,\psi}$, which appears in the surface
gravity (\ref{chi2}), is constant. From Eq.(\ref{Hevenodd}) we find that
\begin{equation}
    3H(1,\psi)+\tan\psi H_{,\psi}(1,\psi)=3\delta_{0}
    +\sin\psi
        -\sum_{k=1}^{\infty}a(\lambda=2k+1)
    \frac{3\sin(2k+1)\psi}{2(2k+1)\sin\psi}\equiv3H(1,\pi/2)~.
    \label{isconstant}
\end{equation}
To solve for $a(2k+1)$, one can multiply (\ref{isconstant})
by $\sin\psi$ and expand in the set
$\left\{\sin(2k+1)\psi\right\}_{k=0,1\ldots}$,
which is complete
and orthogonal over the interval $[0,\pi/2]$ provided that the boundary
conditions are $f(0)=0$ and $f'(\pi/2)=0$.
The coefficients are
\begin{eqnarray}
    & &\delta_{0}=H(1,\pi/2)-\frac{8}{9\pi}~,
    \label{delta0}\\
    & &a(\lambda=2k+1)_{k\geq1}= \frac{-16}{3\pi(\lambda^{2}-4)}
    \label{aodd}~.
\end{eqnarray}
Since we have multiplied (\ref{isconstant})
with $\sin\psi$ one should verify that
(\ref{isconstant}) holds for $\psi=0$, and indeed it is.

At this stage we have to verify the consistency of;
(i) the representation of the function $H$, Eq.(\ref{Hevenodd}),
(ii) the zeroth law of thermodynamics, Eqs.(\ref{delta0}, \ref{aodd}),
and (iii) the first law of thermodynamics, Eq.(\ref{T2T1}).

We use
the following expansions for the Legendre functions around $\rho=1$
\begin{eqnarray}
    \rho\sqrt{\rho^{2}-1}P^{1}_{\nu}(2\rho^{2}-1)&=&\nu(\nu+1)(\rho^{2}-1)
    +\frac{\nu^{2}(\nu+1)^{2}}{2}(\rho^{2}-1)^{2}+{\cal O}(\rho^{2}-1)^{3}~,
    \label{P1}\\
    \rho\sqrt{\rho^{2}-1}Q^{1}_{\nu}(2\rho^{2}-1)&=&-\frac{1}{2}
    +\frac{\nu(\nu+1)}{2}(\rho^{2}-1)\left[1-\ln(\rho^{2}-1)
    -2\gamma_{E}-2\bar{\psi}(\nu+1)\right]\nonumber\\
    & +&\frac{\nu^{2}(\nu+1)^{2}}{4}(\rho^{2}-1)^{2}
    \left[\frac{5\nu^{2}+5\nu+2}{2\nu(\nu+1)}-\ln(\rho^{2}-1)
    -2\gamma_{E}-2\bar{\psi}(\nu+1)\right]
    +{\cal O}(\rho^{2}-1)^{3}~,
    \label{Q1}
\end{eqnarray}
where $\gamma_{E}$ is Euler's constant and $\bar{\psi}(z)$ is the logarithmic
derivative of the Gamma function, $\bar{\psi}(z)=d/dz \ln\Gamma(z)$.

First, we evaluate Eq.(\ref{Hevenodd}) at $\rho=1$,
$\psi=\pi/2$
\begin{eqnarray}
    H(1,\pi/2)&=&\delta_{0}+\frac{1}{8}+\frac{1}{32}-\frac{d_{2}}{2}
    +\frac{3}{32}-\frac{1}{2}
    \sum_{\lambda=3}^{\infty}\frac{(-1)^{\lambda}16}{3\pi(\lambda^{2}-4)}
    \frac{3\sin(\lambda\pi/2)}{\lambda(4-\lambda^{2})}\nonumber\\
    &=&\delta_{0}+\frac{1}{4}-\frac{d_{2}}{2}-\frac{8}{\pi}
    \sum_{k=1}^{\infty}\frac{(-1)^{k}}{(2k+1)((2k+1)^{2}-4)^{2}}
    =\delta_{0}+\frac{8}{9\pi}-\frac{d_{2}}{2}~.
    \label{H1brane}
\end{eqnarray}
Comparing Eq.(\ref{H1brane}) with the result of the first law, Eq.(\ref{delta0}),
we conclude that $d_{2}=0$.

Second, we would like to evaluate the first law, Eq.(\ref{T2T1}).
For this we need to evaluate $H_{,\rho}(1,\psi)$.
We use the expansions (\ref{P1}, \ref{Q1}) in Eq.(\ref{Hevenodd})
and find that
\begin{eqnarray}
    H_{,\rho}(1,\psi)&=&\frac{h_{0}(\psi)}{\sin^3\psi}\left[
    \ln(\rho^{2}-1)+2\gamma_{E}+2\bar{\psi}(1/2)\right]+h_{1}(\psi)~.
    \label{dH1}\\
    h_{0}(\psi)&=&\frac{-7\psi+4\psi\cos2\psi+3\psi\cos4\psi-2\sin2\psi+\sin4\psi}
        {32\pi}\nonumber\\
        & &+\frac{2}{\pi}\sum_{\lambda=3}^{\infty}\frac{(-1)^{\lambda}(\lambda^{2}-1)}
        {\lambda(\lambda^{2}-4)^{2}}\left[\lambda\cos\lambda\psi\sin2\psi
        -2\cos2\psi\sin\lambda\psi\right]~.\label{h0}\\
    h_{1}(\psi)&=&\frac{5\psi-2\sin4\psi+3\psi\cos4\psi}
        {8\pi\sin^3\psi}\nonumber\\
        & &+\frac{4}{\pi}\sum_{\lambda=3}^{\infty}\frac{(-1)^{\lambda}(\lambda^{2}-1)}
        {\lambda(\lambda^{2}-4)^{2}}\left[\bar{\psi}\left(\frac{\lambda+1}{2}\right)
            -\bar{\psi}\left(\frac{1}{2}\right)\right]
        \frac{\lambda\cos\lambda\psi\sin2\psi
        -2\cos2\psi\sin\lambda\psi}{\sin^3\psi}~.\label{h1}
\end{eqnarray}
Let us deal with the function $h_{0}(\psi)$, which multiplies the
diverging term $\ln(\rho^{2}-1)$.
One can verify that $h_{0}(\psi=\pi/2)=0$ and $h_{0}(\psi=0)=0$. Therefore, one can
expand the function $h_{0}(\psi)$
in the set $\left\{\sin2n\psi\right\}_{n=1,2...}$, which is complete
and orthogonal over the interval $[0,\pi/2]$.
It is easy to verify that for $n\geq1$
\begin{equation}
    \int_{0}^{\pi/2} h_{0}(\psi)\sin2n\psi\,d\psi=0~.
\end{equation}
So, the function $h_{0}(\psi)=0$ and
the first derivative of the function $H$ is finite on the
horizon. The first law, Eq.(\ref{T2T1}),
together with Eq.(\ref{delta0}) can be used to find the value of $\delta_{0}$
\begin{equation}
    \delta_{0}=\frac{8}{27\pi}\left[-2+h_{1}'(\pi/2)
    +\int_{0}^{\pi/2}\frac{d\psi}{\cos\psi}\left[
    h_{1}(\psi)\sin^{3}\psi \right]_{,\psi}\right]~.
    \label{firstlaw}
\end{equation}
From Eq.(\ref{h1}) one can deduce the following
\begin{eqnarray}
    & &\left[h_{1}(\psi)\sin^{3}\psi\right]_{,\psi}=
    \frac{5-5\cos4\psi-12\psi\sin4\psi}
        {8\pi}-\frac{4}{\pi}\sum_{\lambda=3}^{\infty}\frac{(-1)^{\lambda}(\lambda^{2}-1)}
        {\lambda(\lambda^{2}-4)}\left[\bar{\psi}\left(\frac{\lambda+1}{2}\right)
            -\bar{\psi}\left(\frac{1}{2}\right)\right]\sin2\psi\sin\lambda\psi~.\\
    & &h_{1}'(\pi/2)=\left.\left[h_{1}(\psi)\sin^{3}\psi\right]_{,\psi}
    \right|_{\psi=\pi/2}=0~.\\
    & &\int_{0}^{\pi/2}\frac{d\psi}{\cos\psi}\left[
    h_{1}(\psi)\sin^{3}\psi\right]_{,\psi}=\frac{5}{\pi}+
    \frac{2}{\pi}\sum_{n=2}^{\infty}\frac{(-1)^{n}}
        {(n^{2}-1)}\left[\bar{\psi}\left(n+1/2\right)
            -\bar{\psi}\left(1/2\right)\right]=2~.
\end{eqnarray}
Now, substitute in Eq.(\ref{firstlaw}) to find $\delta_{0}=0$.

To summarize, if we give up the post linearized assumption and allow for
asymptotic behavior of $M^{n/2}$ then we can satisfy the requirement that
the surface gravity is constant and the solution appears to be regular
on the horizon. The function $H$ and the first derivative $H_{,\rho}$ are
finite on the horizon. The derivatives $H_{,\psi}$ and $H_{,\psi\psi}$ are
vanishing there. Therefore, based on Eq.(\ref{Hequation}) we can conclude
that $H_{,\rho\rho}$ is finite on the horizon as well. This is a sufficient condition
for the metric to be $C^{1}$. In general this is not sufficient to ensure that
the the metric is not singular.
To double check, we look at the Kretchmann scalar up to first order in $\epsilon$
\begin{equation}
    K=R^{abcd}R_{abcd}=\frac{72}{\rho^{8}}-\epsilon\left[
    \frac{576}{\rho^{8}}\left(\frac{}{}F(\rho,\psi)
    +\rho(\rho^{2}-1)\cos\psi\right)+\frac{96}{\rho}
    \left(\frac{3H(\rho,\psi)+\tan\psi H_{,\psi}(\rho,\psi)}
    {\rho^{6}}\right)_{,\rho}\right]~.
    \label{Kretch1}
\end{equation}
As one can see, the Kretchmann scalar depends only on
the first derivative $H_{,\rho}(\rho,\psi)$, and therefore it is finite.
An ambiguity is left for the second order in $\epsilon$. The
Kretchmann scalar at the second order in $\epsilon$ depends on
higher derivatives of $H$, which might be divergent.
However,
second order in $\epsilon$ in the Kretchmann scalar depends
on the first and second orders in the metric. So nothing can be
done at this stage before finding the second order corrections to
the metric. These should cancel the divergence part coming from
the first order in the metric.
We should emphasize that the situation with the surface
gravity is different. It is true that
if the surface gravity is not constant in
first order in $\epsilon$, it will affect the Kretchmann scalar
only at second order in $\epsilon$. However, this contribution
cannot be compensated by second order corrections to the metric.
Therefore, the surface gravity must be constant at all orders.

What about the asymptotic behavior of the metric?
The inclusion of half integer powers of the mass changes the post
Newtonian potential. For large $\rho$, the terms with
$a(\lambda=2k+1)_{k\geq1}$ contribute to the function $H$ (and
throughout Eq.(\ref{Fbrane}) to the function $F$) terms of order
$\rho^{-2k}$. As a result, the gravitational potential, $g_{tt}$,
acquires terms of the form
\begin{equation}
    \epsilon \frac{a(\lambda=2k+1)}{\rho^{2k+2}}\sim
    \frac{a(\lambda=2k+1)(G_{5}M)^{k+3/2}}{\bar{\rho}^{2k+2}\ell}~,
\end{equation}
where $\bar{\rho}=\mu\rho$ is a dimensionful coordinate. The term
with $\delta_{0}$ would have contributed a term of the form
$\delta_{0}(G_{5}M)^{3/2}\bar{\rho}^{-2}\ell^{-1}$. But, since
$\delta_{0}=0$ the half integer terms start only from $M^{5/2}$.

We learn that, in general, the asymptotic solution should be
expanded in powers of $(G_{5}M)^{1/2}$. The lowest order must be
$(G_{5}M)^{1}$ and not $(G_{5}M)^{1/2}$ otherwise there will not
be a well defined conserved mass. One might expect that by
iterating the lowest order there will be only integer powers of
$M$. However, the boundary conditions on the horizon require also
half integer powers of the mass. In particular it is the zeroth
law of black hole thermodynamics (constant surface gravity) that
forced us to include the half integer terms. The coefficients of
the terms $(G_{5}M)^{k/2},\;k=5,7\ldots$, which are
$a(\lambda=2k+1)_{k\geq1}$, are completely determined by the
zeroth law (\ref{aodd}). The coefficient of $(G_{5}M)^{3/2}$,
which is $\delta_{0}$, is not fixed by the requirement that the
surface gravity is constant. The actual value of the surface
gravity depends on $\delta_{0}$. The first law of black hole
thermodynamics, $dM=T\,dS$, forces $\delta_{0}$ to vanish, and
therefore eliminates the term of order $(G_{5}M)^{3/2}$ from the
asymptotic expansion.

The dominant part of the post Newtonian
potential remains
$(G_{5}M)^{2}/(r^{2}\ell^{2})$, which is responsible for the
precession of perihelion calculations.
Since we don't know of any experimental evidence for higher
order terms in the potential, it looks like the half
integer terms cannot be detected experimentally.

\section{Summary and Discussion}
We study a small black hole located on the TeV brane in Randall-Sundrum I
(two branes) scenario.
We expand the metric in $\epsilon$, which is the ratio between the
Schwarzschild radius and the bulk curvature length.
We find the solution up to first order in $\epsilon$.
The solution satisfies Einstein's equations in the bulk and
Israel junction conditions on the TeV brane.
The asymptotic form of the metric is fixed by the weak field
approximation (linearized gravity).

In Randall-Sundrum II (single brane scenario)\cite{rs2} it was
conjectured that there were no large static black
holes localized on the brane\cite{noBH}.
However, small black holes might exist and one
can apply our method of $\epsilon$ expansion to find them. Although the
linearized solution for a single brane is very different from that
for two branes, in first order of $\epsilon$ they are similar.
The only difference is $\epsilon\rightarrow
-\epsilon$. So, the results for the $\epsilon$ solution and the
discussion of regularity are equivalent in the two scenarios.

A crucial issue is the post linearized form of the asymptotic
solution.
For Four dimensional black holes the post Newtonian behavior
is given in terms of an expansion
in integer powers of $G_{4}M/r$ (the only dimensionless combination).
The equivalent in RS models would be
an expansion in integer powers of $G_{5}M/(r\ell)$.
However, if we assume that the
post linearized metric includes only integer powers of the mass
then the surface gravity on the horizon is not constant. This
means that the horizon is singular, and the configuration of a
small mass on the TeV brane describes a naked singularity.
As we mentioned
earlier, the singularity is naked from the bulk but it is covered
on the brane.

In section~\ref{sec:postlinearized} we have shown that there is a
solution with a regular horizon, i.e. the surface gravity is
constant and the Kretchmann scalar is finite up to first order in
$\epsilon$. However, this requires a different asymptotic form of
the metric, as one must include half integer powers of the mass.
In general, the expansion parameter for the asymptotic solution
is $(G_{5}M)^{1/2}$, but still the lowest order is $(G_{5}M)^{1}$ and not
$(G_{5}M)^{1/2}$ otherwise there will not be a well defined
conserved mass. Surprisingly, the first law of black hole
thermodynamics, $dM=T\,dS$, eliminates the term of order
$(G_{5}M)^{3/2}$, as well, from the asymptotic expansion. The leading term
in the post Newtonian potential remains $(G_{5}M)^{2}/(r^{2}\ell^{2})$,
which is the term responsible for the precession of
perihelion calculations in the Schwarzschild metric.
The next order term is $(G_{5}M)^{5/2}$. This term does not exist
in the four dimensional Schwarzschild metric, but it is very hard
to detect such a term in any asymptotic measurement.

As we mentioned earlier, these results are also valid for small black holes
in RS single brane
scenario. In principle, the post Newtonian behavior can be verified numerically.
However, the asymptotic behavior of the numerical solution of \cite{kudoh}
is not well known \cite{email}. Future numerical studies with a regular horizon
might detect the fractional power behavior in the post Newtonian potential.

The configuration discussed in this paper, of a small (few TeV)
mass on the brane in RS scenario, can be the final state of high
energy particle collisions. In some sense such a collision is
similar to the collapse of matter in ordinary four dimensional
gravity, it might end up as a black hole or generate a naked
singularity\cite{cenalo}. If the mass of the object is somewhat
bigger than the Planck mass (TeV) the behavior of these two
possibilities is very different. The black hole will radiate
thermally. It will mainly radiate ordinary particles on the brane.
The naked singularity has no temperature, it evaporates in an
explosive fashion in the order of Planck time. It will mainly
radiate gravitons into the bulk (where the singularity is naked)
and not on the brane (where it is covered). further investigations
of this issue will be carried out elsewhere.

\begin{acknowledgments}
This work is supported in part by the U.S. Department of Energy Grant
No. DE-FG02-84ER40153.
We thank Richard Gass and Cenalo Vaz for fruitful discussions.
\end{acknowledgments}

\end{document}